  \documentclass{cjour}
\def\half{\mbox{$\frac{1}{2}$}}     % small built-up `one-half'
\def\eqcomma{\quad ,}
\def\eqperiod{\quad .}
\def\bY{\bar{Y}}                    % complex conjugate of Y
\def\bra{\langle}
\def\ket{\rangle}

\begin{document}

%%%%% To be entered at Academic Press: =====>>

% \journame{J Comp Phys}
% \articlenumber{}
% \yearofpublication{}
% \volume{}
% \cccline{}
 \received{12 October 1999}
% \revised{}
% \accepted{}

\authorrunninghead{C. W. Misner}
\titlerunninghead{$Y_l^m$ on a Cubic Grid}

%% <<== End of commands to be entered at Academic Press

%%  Authors, start here ==>>

%\draft % Optional, will cause a line at the bottom of each page
%% with the words `Draft' and the time and date that the article
%% was LaTeXed. Will also double space text.

\title{Spherical Harmonic Decomposition\\
            on a Cubic Grid%
%\protect\cite{PP}
\thanks{UM PP-00-032; gr-qc/9910044}
}

\author{Charles W. Misner}
\affil{
            Department of Physics, University of Maryland,
            College Park MD 20742-4111 USA
}

\email{misner@umail.umd.edu}

%\date{12 October 1999}

\abstract{
A method is described by which a function defined on a cubic grid 
(as from a finite difference solution of a partial differential
equation) can be resolved into spherical harmonic components at
some fixed radius.  
This has applications to the treatment of boundary conditions 
imposed at radii larger than the size of the grid, following 
Abrahams, Rezzola, Rupright et al.\ \cite{arr97}.  
In the method described here, the interpolation of the grid data 
to the integration 2-sphere is combined in the same step as the
integrations to extract the spherical harmonic amplitudes, which
become sums over grid points.
Coordinates adapted to the integration sphere are not needed.
}


\begin{article}

\section{Introduction}
\label{sec:Introduction}

Abrahams, Rezzola, Rupright, et al.\ \cite{arr97,rar98,ram98} 
used spherical harmonics to interpolate between a 3D grid on
which sources of gravitational radiation were modeled and a 
distant boundary on which outgoing wave boundary conditions were
imposed.

While applying this method to a simplified model problem 
(scalar gravity \cite{kwcm99}), we have developed an accurate 
and efficient method for extracting spherical harmonic amplitudes 
from the data on a rectangular grid.
Spherical harmonic amplitudes are in principle defined as
\begin{equation}
\label{eq:SphHarmDefn}
\Phi_{lm}(r,t) = \oint \bY^{m}_{l}(\theta,\phi)
        \Phi(r,\theta,\phi,t) \,d^2 \Omega
\end{equation}
by integrating over a sphere of radius r.  Here $\Phi$ is the scalar 
field whose spherical harmonic decomposition
\begin{equation}
\label{eq:PhiDecomp}
\Phi(r,\theta,\phi,t) =\sum_{lm}\Phi_{lm}(r,t) Y^{m}_{l}(\theta,\phi)
\end{equation}
is desired.

Our method combines in one step the interpolation from the 
rectangular grid to the sphere and the integration over the sphere.
It is motivated by the remark that a surface integral is the derivative 
of a volume integral
\begin{equation}
\label{eq:DVolIntegral}
\Phi_{lm}(r,t) = \frac{d}{dr} \int_{0}^{r} \!\int \!\! \int \bY^{m}_{l} 
\Phi(r',\theta,\phi,t) \,d^2 \Omega \,dr'
\eqperiod
\end{equation}
Volume integrals are relatively straightforward on a rectangular grid.
A finite difference approximation to (\ref{eq:DVolIntegral}) gives an
integral over a shell of finite thickness covering the sphere $r=$
constant. 
Our full implementation described below has some additional 
features.

\section{Continuum Theory}
\label{sec:ContinuumTheory}
  
In order to find spherical harmonic amplitudes $\Phi_{lm}(R,t)$ at a
given fixed radius $R$, and also their radial derivatives 
$\partial_r \Phi_{lm}(r)]_{r=R}$, we first consider a complete 3D 
``Fourier" decomposition of $\Phi(x,y,z,t)$ in a shell of thickness 
$2\Delta$ around $R$.
As a set of orthonormal functions on this shell we take the 
functions
\begin{equation}
\label{eq:OrthoFunctions}
Y_{nlm}(r,\theta,\phi) = R_n(r) Y^{m}_{l}(\theta, \phi)
\eqcomma
\end{equation}
where the $R_n$ are obtained from Legendre polynomials,
\begin{equation}
\label{eq:LegendrePolys}
R_n(r) = P_n(\frac{r-R}{\Delta}) 
\frac{1}{r}\sqrt{\frac{2n+1}{2\Delta}}
\eqcomma
\end{equation}
which have the property
\begin{equation}
\label{eq:ROrtho}
\int^{R+\Delta}_{R-\Delta} R_n(r) R_k(r) r^2 dr = \delta_{nk}
\end{equation}
as a consequence of the orthogonality of the Legendre polynomials
\begin{equation}
\label{eq:LegendreOrtho}
\int^{1}_{-1} P_n(x) P_k(x) \,dx = 2\delta_{nk} / (2n + 1)
\eqperiod
\end{equation}
Then the functions $Y_{nlm}$ are a complete orthonormal set of 
functions on the shell $S = \{R-\Delta \leq r \leq R+\Delta\}$ using
the inner product
\begin{equation}
\label{eq:InnerProduct}
\bra f | g \ket = \int \!\!\int \!\!\int_{S} \bar{f}(x,y,z) 
                      g(x,y,z) \,dx \,dy \,dz
\eqperiod
\end{equation}
and its associated norm $\parallel f \parallel^2 \equiv \bra f | f
\ket$.
When a function $\Phi(x,y,z)$ is smooth enough that the 
generalized Fourier series converges pointwise on the sphere 
$r=R$, one has
\begin{equation}
\label{eq:PhiOnShell}
\Phi(R,\theta,\phi,t) =\sum_{nlm}\Phi_{nlm} R_n(R) 
Y^{m}_{l}(\theta \phi)
\end{equation}
so that the spherical harmonic amplitudes for $\Phi$ at $r=R$, 
i.e., the coefficients of $Y^{m}_{l}$, are
\begin{equation}
\label{eq:PhiCoefficients}
\Phi_{lm}(R) =\sum_{n}\Phi_{nlm} R_n(R) 
\eqperiod
\end{equation}
 Since $R_n(R) = 0$ for odd $n$, this sum in numerical applications
 will typically have only one or two terms, {\it e.g.}, $n=0$ and 
 possibly $n=2$.
 
The coefficients $\Phi_{nlm}$ are calculated from the volume 
integrals that are evaluated numerically on the rectangular grid:
\begin{equation}
\label{eq:Phinlm}
\Phi_{nlm} = \int \!\!\int \!\!\int_{S}  \bY^{}_{nlm}(r,\theta,\phi) 
\Phi(r,\theta,\phi) \,dx \,dy \,dz
\end{equation}
so
\begin{equation}
\label{eq:Philm}
\Phi_{lm}(R) = \int \!\!\int \!\!\int_{S} \left[\sum_n R_n(R) R_{n}(r)
             \right]   
      \bY^{m}_{l}(\theta,\phi) \Phi(r,\theta,\phi) 
      \,dx \,dy \,dz
\eqperiod
\end{equation}

\section{Discretization}
\label{sec:Discretization}

When the scalar function $\Phi(x,y,z)$ is known only at points $x$
on a cubic grid $G$ with spacing $\Delta x = \Delta y = \Delta z =
k$, the inner product (\ref{eq:InnerProduct}) for our generalized
Fourier analysis is relaced by
\begin{equation}
\label{eq:DiscreteInnerProduct}
\bra f | g \ket = \sum_{x \in G} \bar{f}(x) g(x) \,w_x
\eqcomma
\end{equation}
where each grid point must be given some weight $w_x$.
If a cube of volume $k^3$ centered on the grid point lies entirely
outside the shell $S$, the weight will be zero.
If it lies entrely inside $S$, the weight will be its volume $k^3$.
For other grid points we found it overly complicated to calculate,
for the associated cube, the volume which overlaps the shell
$R-\Delta \leq r \leq R+\Delta$.
We simply use the result for a grid point lying on one of the 
rectangular axes ({\it i.e.}, we imagined the cube rotated so its
edges were parallel or perpendicular to the radial direction).
Thus we took
\begin{equation}
\label{eq:WeightingFunction}
w_r = 
\left\{ \begin{array}{ll}      
        0 \quad &\mbox{if $\quad |r-R| > \Delta + \half k$} \\
        k^3 \quad &\mbox{if $\quad |r-R| < \Delta - \half k$} \\
       (\Delta+\half k - |R-r|)k^2 &\mbox{otherwise}
       \end{array}
   \right.
\end{equation}
where $r$ is the Euclidean distance from the origin, and considered 
only the case $k < 2 \Delta$.

With the inner product modified by this discretization, the function 
set (\ref{eq:OrthoFunctions}) will no longer be orthonormal.
One will find that
\begin{equation}
\label{eq:NormalizationTensor}
\bra Y_A | Y_B \ket = G_{AB} = \bar{G}_{BA}
\end{equation}
is no longer a unit matrix.  Here $A$ and $B$ are index groups 
$(nlm)$ so that $Y_A$ abbreviates the $Y_{nlm}(r,\theta,\phi)$
of (\ref{eq:OrthoFunctions}).
The mean square error (as measured by this new norm 
\ref{eq:DiscreteInnerProduct}) in the representation
\begin{equation}
\label{eq:PhiApproximation}
\Phi(x,y,z) \approx \sum_{A} \Phi^{A} Y_{A}(x,y,z)
\end{equation}
is now minimized by the choice
\begin{equation}
\label{eq:PhiChoice}
\Phi^{A}  = \sum_{B} G^{AB} \bra Y_B | \Phi \ket
\end{equation}
where $G^{AB}$ is the matrix inverse to $G_{AB}$.
This holds for any finite subset of the $Y_{nlm}$ one chooses to 
employ.  Note that if one defines ``adjoint'' harmonics by
\begin{equation}
\label{eq:PhiAdjoint}
    Y^{A}  = \sum_{B} Y_B G^{BA}
\end{equation}
then these coefficients in (\ref{eq:PhiApproximation}) can 
be computed as
\begin{equation}
\label{eq:PhiCoef}
\Phi^{A}  =  \bra Y^A | \Phi \ket
\eqperiod
\end{equation}
Similarly, by defining
\begin{equation}
\label{eq:YlmCoef}
  R_{lm}  = \sum_n \bar{R}_n(R) Y^{nlm}(r,\theta,\phi)
\end{equation}
in which the adjoint $Y^A$ appear, not the original $Y_A$ of
equation (\ref{eq:OrthoFunctions}), one calculates directly 
the spherical harmonic amplitudes of 
equation~(\ref{eq:PhiCoefficients}) by evaluating the sum
\begin{equation}
\label{eq:PhiCoeff}
\Phi_{lm}  =  \bra R_{lm} | \Phi \ket
\eqperiod
\end{equation}

It is important to notice that, in typical applications following
the model of \cite{arr97}, the grid and the sphere $r=R$ at which 
a spherical harmonic decomposition is used do not change while
the function $\Phi$ or other fields are dynamically evolved or 
are relaxed toward a stationary state.  Then a possible
strategy is to store in memory the quantities $R_{lm}(x) w_x$ at 
each grid point within the integration shell so that the
computation of spherical harmonic coefficients becomes
\begin{equation}
\label{eq:DiscreteCoef}
  \Phi_{lm}(R) = \sum_{x \in G} \bar{R}_{lm}(x) \,w_x \Phi(x)
\end{equation}
where only $\Phi$ changes at each evolution or relaxation step
and numerous other complications such as the $G_{AB}$, the
$R_n(r)$, and the $w_x$ are buried in a large table which does
not change at each iteration of the p.d.e., nor even at each run
of the program.

\section{Example}
\label{sec:Example}

As a test to illustrate the application of these ideas, Mr. Keith
Watt has run a test of the recovery of known spherical harmonic
coefficients by these means.  A cubic grid has each $x,y,z$
coordinate varying at intervals of $k=0.2$ in the
range $-1.3 \leq x,y,z \leq +1.3$. 
A known function $\Phi$ is evaluated at these $14^3 = 2744$ 
grid points.  
The sphere on which the resolution into spherical harmonics 
was desired was specified by $R=1.0$ 
and the thickness of the shell approximating this sphere was 
fixed by $\Delta = 0.15$ in the notations of 
equation~(\ref{eq:WeightingFunction}).  
Thus nonzero weight $w_g$ was assigned to grid points $g$ at 
radii in the range $0.75 < r < 1.25$, which includes about 
800 points.  
The approximately 1720 points at $r > 1.25$ and the 
approximately 220 points at $r<0.75$ thus play no role in 
this anaylsis.

The choice $\Delta = (3/4)k$ was motivated by the fact that, in a
one dimensional analogue of this analysis, this choice leads to
Simpson's rule for the numerical integration and diagonalizes the
$3\times3$ matrix $G_{AB}$ when one uses only $n \leq 2$.

    The test problem used 
\begin{equation}
\label{eq:testPhi}
   \Phi = \sum \Phi_{lm} (r/R)^l Y_l^m
\end{equation}
with $0 \leq l \leq 2$ for a total of nine terms using real
valued spherical harmonics.  
The input values $\Phi_{lm}$ and the recovered values $B_{lm}$ 
are shown in the table.  
The accuracy was less (0.1\% rather than
the typical 0.01\% shown in the table) when the radial dependence
was changed to $(R/r)^{l+1}$; but the function used for the table
may be more like the behaviors expected in applications where the
sphere in question is outside the strong source region but within
a wavelength of the center.

\begin{table}
\begin{tabular}{lrrrr}

$\ell$   &    m & $\Phi_{lm}$ &   $B_{lm}$     &    \% error
\\ \hline
0   &    0    &   9.000000    &    8.999315    &   -0.0076  \\
1   &    -1   &   8.000000    &    8.000664    &    0.0083  \\
1   &    0    &   7.000000    &    6.999812    &   -0.0027  \\
1   &    1    &   6.000000    &    5.998907    &   -0.0182  \\
2   &    -2   &   5.000000    &    5.000006    &    0.0001  \\
2   &     -1  &   4.000000    &    4.000457    &    0.0114  \\
2   &    0    &   3.000000    &    3.001447    &    0.0482  \\
2   &    1    &   2.000000    &    2.000002    &    0.0001  \\
2   &    2    &   1.000000    &    0.999975    &   -0.0025  
\end{tabular}
\caption{Spherical harmonic amplitudes:   
          $\Phi$ input, $B$ recovered. }
\end{table}
%% End of article:

\begin{acknowledgment}
I thank Conrad Schiff and Keith Watt for many helpful conversations.  Keith Watt  
performed the computations to test this method.  I thank Joan Centrella 
and Richard Matzner for valuable discussions in the early stages of 
formulating a research program from which this problem arose. 
This research was supported in part by NSF grant PHY 9700672.
\end{acknowledgment}

%% This command is necessary! ==>>
\end{article}
\end{document}